\documentstyle[prl,aps,twocolumn,epsfig,amstex]{revtex}

\newcommand{\nn}{\nonumber\\} 
 
\newcommand{\f}[1]{\mbox{\boldmath$#1$}}

\newcommand{\na}{\mbox{\boldmath$\nabla$}}
\newcommand{\bea}{\begin{eqnarray}}
\newcommand{\ea}{\end{eqnarray}}

\begin{document} 
 
\wideabs{
\title{Hawking radiation in an electro-magnetic wave-guide?}
\author{Ralf Sch\"utzhold$^{1}$ and William G.~Unruh$^{2,3}$}
\address{
$^1$Institut f\"ur Theoretische Physik,
Technische Universit\"at Dresden,
01062 Dresden, Germany
\\
$^2$Department of Physics and Astronomy,
University of British Columbia,
Vancouver B.C., V6T 1Z1 Canada
\\
$^3$Canadian Institute for Advanced Research Cosmology and Gravity Program
\\
Electronic addresses: 
{\tt schuetz@@theory.phy.tu-dresden.de};
{\tt unruh@@physics.ubc.ca}
}
\date{\today}
\maketitle
\begin{abstract} 
It is demonstrated that the propagation of electro-magnetic waves in
an appropriately designed wave-guide is (for large wave-lengths)
analogous to that within a curved space-time -- 
such as around a black hole.  
As electro-magnetic radiation (e.g., micro-waves) can be controlled, 
amplified, and detected (with present-day technology) much easier
than sound, for example, we propose a set-up for the experimental
verification of the Hawking effect.
Apart from experimentally testing this striking prediction, this would
facilitate the investigation of the trans-Planckian problem.
\end{abstract} 
}

{\em Introduction}\quad
%
One of the major motivations behind the idea of black hole analogues 
(''dumb holes'', see \cite{unruh-prl}) is the possibility of an 
experimental verification of the Hawking effect \cite{hawking}. 
Apart from testing one of the most striking theoretical predictions 
of quantum field theory under the influence of external conditions, 
such an experiment would enable us to investigate the impact of 
ultra-high energy/momentum degrees of freedom (trans-Planckian problem)
on the lowest-order Hawking effect and its higher-order corrections 
(with respect to the small ratio of Hawking temperature over Planck scale) 
by means of an analogue system.
In view of the close relation between the Hawking effect and the
concept of black hole entropy, these investigations are potentially
relevant for the black hole information paradox etc.

The analogy between sound waves in moving fluids and scalar fields in
curved space-times established in \cite{unruh-prl} can (in principle)
be used to simulate a horizon in liquid Helium \cite{helium} 
or in Bose-Einstein condensates \cite{bec}, for example
(see also \cite{artificial}).
However, measuring the Hawking effect in those systems goes along with
serious difficulties \cite{pessi}. 
The main problem is the detection of sound waves corresponding to the
realistically very low Hawking temperature.

On the other hand, electro-magnetic radiation is much easier to
amplify and to detect with present-day technology and one might hope  
to exploit this advantage.
Optical black hole analogues have been discussed for highly dispersive
media which support the phenomenon of slow light \cite{slow}  
and for ordinary non-dispersive dielectrics \cite{dielectric}.
However, it turns out that a horizon for slow light does not emit
Hawking radiation \cite{not-slow} whereas an experimental realization
by means of ordinary dielectrics is in principle possible but very
challenging. 
In the following, we shall propose an alternative set-up in order to
circumvent this difficulty.

\begin{figure}[ht]
\centerline{\mbox{\epsfxsize=8.5cm\epsffile{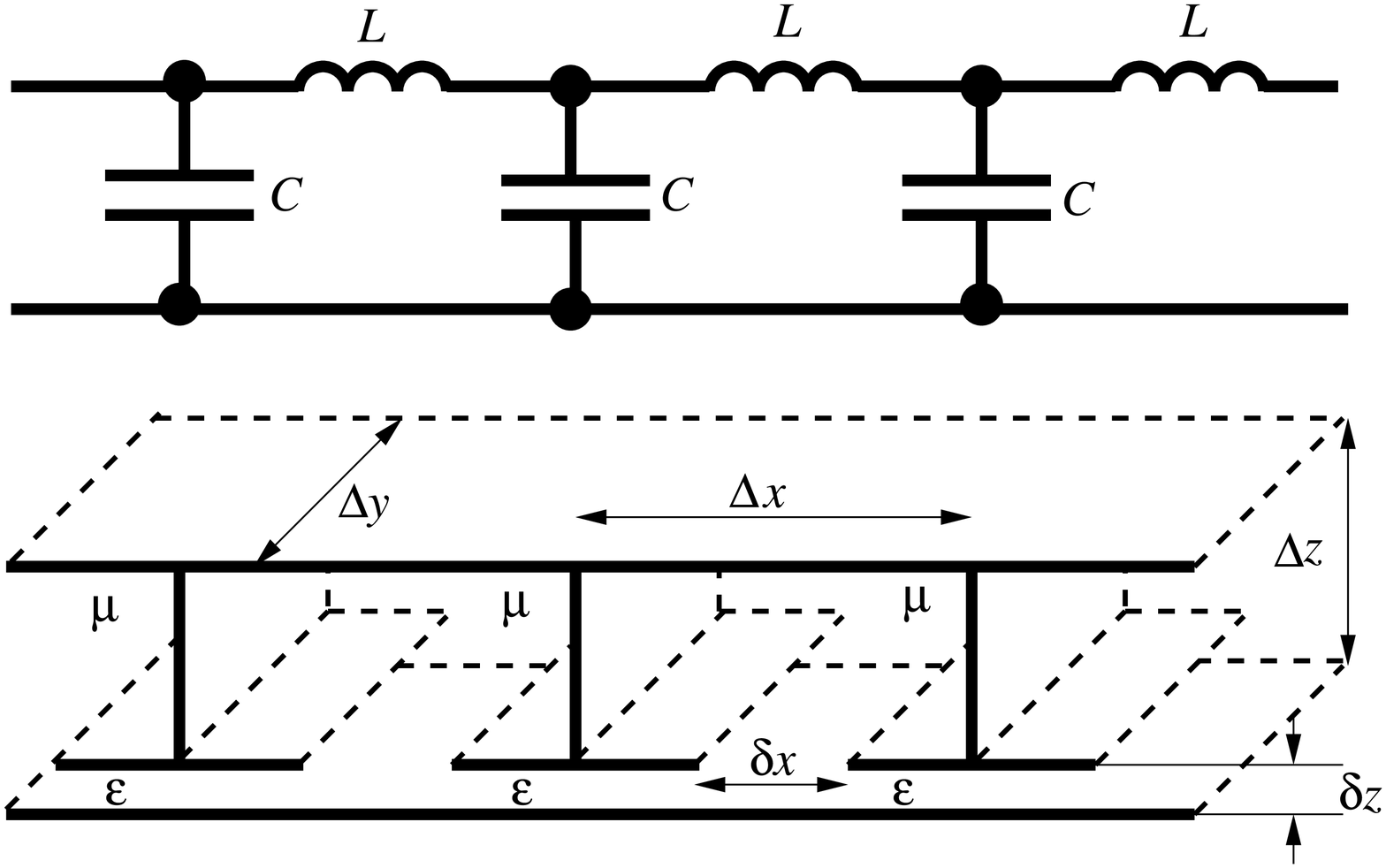}}}
\caption{Circuit diagram and sketch of wave-guide. \\
The capacitances $C$ are realized by the parallel conducting
plates at the bottom of each unit separated by insulating slabs with 
the dielectric permittivity $\varepsilon$ and thickness $\delta z$.
The inductances $L$ are generated by the remaining space of the 
$\Delta x\times\Delta y\times\Delta z$ cells 
(magnetic permeability $\mu$) with the enclosing walls also
being conductors.}   
\label{wellen}
\end{figure}

{\em Wave-guide}\quad
%
Let us consider the propagation of electro-magnetic waves in the
wave-guide in Figure~\ref{wellen}.
For simplicity, and in order to avoid leakage and transverse modes etc., 
we assume the following hierarchy of dimensions of the wave-guide 
and the wave-length $\lambda$ of the propagating electro-magnetic waves
\bea
\label{delta}
\delta z \ll \delta x \ll \Delta x,\Delta z \ll \Delta y \ll \lambda
\,.
\ea
In this limit, the wave-guide possesses a large slow-down and we can
omit Maxwell's supplement $\f{\dot D}$ in Oerstedt-Amp\`ere's law 
$\na\times\f{H}=\f{j}+\f{\dot D}\;\leadsto\;
\oint d\f{r}\cdot\f{H}=I+\frac{d}{dt}\int d\f{S}\cdot\f{D}$
in the upper region of the wave-guide (i.e., the surface integral over 
$S=\Delta x\times\Delta z$).
The conditions (\ref{delta}) also ensure that the energy of the waves
is basically confined to the wave-guide.
Hence, the combination of Oerstedt-Amp\`ere's and Faraday's law
$\na\times\f{E}=-\f{\dot B}\;\leadsto\;
\oint d\f{r}\cdot\f{E}=-\frac{d}{dt}\int d\f{S}\cdot\f{B}$
gives the induction law $\Delta U=\frac{d}{dt}LI$ for the effective 
coils with a possibly time-dependent inductance $L(t)$ given by 
$L=\mu\Delta x\Delta z/\Delta y$ for a long coil 
$\Delta y\gg\Delta x,\Delta z$.
Denoting the voltage impressed on the $n$-th capacitor by $U_n$, 
the above condition implies for the $n$-th coil
$U_{n+1}-U_{n}=\frac{d}{dt}\,L_nI_n$.
By means of Gauss' law 
$\na\cdot\f{D}=\varrho\;\leadsto\;\oint d\f{S}\cdot\f{D}=Q$
and the conditions (\ref{delta}), we obtain for the $n$-th capacitor 
$Q_n=C_nU_n$ with $Q_n$ denoting its charge and 
$C_n=\varepsilon_n\Delta x\Delta y/\delta z$ its possibly time-dependent
capacitance.
In the following \cite{dual} we shall assume that all the inductances
are constant and equal ($L_n=L=\rm const$) whereas the capacitances
are generally time-dependent $C_n(t)$.
In analogy to the electro-magnetic field in vacuum, we may introduce
effective potentials $A_n$ which automatically satisfy the induction
law $U_{n+1}-U_{n}=\frac{d}{dt}\,L_nI_n$ via
\bea
\label{potential}
U_n=L\,\frac{dA_n}{dt}
\quad,\quad
I_n=A_{n+1}-A_n
\,.
\ea
Then Kirchoff's law 
$\frac{d}{dt}\,Q_n=\frac{d}{dt}\,C_nU_n=I_n-I_{n-1}$
(i.e., charge conservation) implies the equation of motion 
\bea
\label{discrete}
\frac{d}{dt}\,LC_n\,\frac{d}{dt}\,A_n=A_{n+1}-2A_n+A_{n-1}
\,.
\ea
%

{\em Effective geometry}\quad
%
In the continuum limit 
(i.e., one wave-length involves many units $\lambda\gg\Delta x$),
the above equation of motion approaches the wave equation
\bea
\label{continuum}
\left(
\frac{\partial}{\partial t}\,\frac{1}{c^2}\,\frac{\partial}{\partial t}
-\frac{\partial^2}{\partial x^2}
\right)A=0
\,,
\ea
with the space-time dependent velocity of propagation 
\bea
\label{propagation}
c=\frac{\Delta x}{\sqrt{LC}}=
\sqrt{\frac{\delta z}{\epsilon\mu\Delta z}}\ll c_0
\,.
\ea
If we arrange $c^2(t,x)$ according to $c^2(t,x)=c^2(x+vt)$
with a constant velocity $v$ and transform into the co-moving frame 
($x \to x+vt$), the wave equation becomes
\bea
\label{co-moving}
\left[
\left(\frac{\partial}{\partial t}+v\,\frac{\partial}{\partial x}\right)
\frac{1}{c^2}
\left(\frac{\partial}{\partial t}+v\,\frac{\partial}{\partial x}\right)
-\frac{\partial^2}{\partial x^2}
\right]A=0
\,.
\ea
Unfortunately, in 1+1 dimensions, the Maxwell equations are trivial
and the scalar field is conformally invariant -- which prevents the
introduction of an effective geometry at this stage 
(i.e., in 1+1 dimensions).
On the other hand, the wave-guide is not really 1+1 dimensional, the 
$y$-dimension just does not contribute due to $\lambda\gg\Delta y$,
see \cite{compact}.
Taking into account this ''silent'' $y$-dimension, the above wave
equation allows for the identification of an effective metric via 
$\Box_{\rm eff}A=\partial_\mu(\sqrt{g_{\rm eff}}\,g^{\mu\nu}_{\rm eff}
\partial_\nu A)=0$ in the 2+1 dimensional
Painlev{\'e}-Gullstrand-Lema{\^\i}tre form \cite{pgl} 
\bea
\label{eff-metric}
g^{\mu\nu}_{\rm eff}=
\left(
\begin{array}{ccc}
1 & v & 0 \\
v & v^2-c^2 & 0 \\
0 & 0 & -c^2
\end{array}
\right)
\,.
\ea
Hence, the propagation of electro-magnetic waves within the wave-guide is 
equivalent to that in a curved space-time as described by the above metric.
Note that the scenario under consideration is similar to the super-sonic 
domain wall discussed in \cite{domain}.

As one would expect, the metric describes a horizon at $v^2=c^2$ with
a surface gravity corresponding to the gradient of the propagation
velocity $c$ (since $v=\rm const$) and, therefore, a Hawking
temperature of
\bea
\label{Hawking}
T_{\rm Hawking}=\frac{\hbar}{2\pi k_{\rm B}}\,
\left|\frac{\partial c}{\partial x}\right|_{v^2=c^2}
\,.
\ea
As demonstrated in \cite{not-slow}, the (classical) wave equation is
not enough for the prediction of Hawking radiation -- the (quantum)
commutation relations have to match as well.
If we start from the effective action of the wave-guide
\bea
\label{eff-action}
{\mathfrak A}_{\rm eff}
&=&
\int dt\sum\limits_n\frac{1}{2}\left(C_n\,U_n^2-L_n\,I_n^2\right)
\nn
&\to&
\frac{L\Delta x}{2}
\int dt\,dx\left(
\frac{1}{c^2}\,\left[\frac{\partial A}{\partial t}\right]^2-
\left[\frac{\partial A}{\partial x}\right]^2
\right)
\,,
\ea
and perform the usual canonical quantization procedure, we indeed
obtain the correct commutation relations for the Hawking effect
-- i.e., the conversion of the zero-point oscillations 
(with energy $\hbar\omega/2$)
to real photons by the external conditions $C_n(t)$.

{\em Switching}\quad
%
In the following we present a microscopic model for the controlled
change of the capacitances $C_n(t)$.
Let us consider an insulating material 
(e.g., a semi-conductor at very low temperatures)
where the electrons are localized and their dynamics 
(under the influence of an electro-magnetic field)
can be described by the usual non-relativistic first-order Lagrangian
density  
\bea
\label{total}
{\cal L}=\Psi^*\left(
i\partial_t+\frac{\na^2}{2m}-V+ie\frac{\f{A}\cdot\na}{m}-e\Phi
\right)\Psi
\,.
\ea
Here $m$ denotes the mass and $e$ the charge of the electrons, 
$V(\f{r})$ is the static one-particle potential, and $\f{A}$ and
$\Phi$ are the vector and scalar potentials of the electro-magnetic
field, respectively.  
The scalar potential $\Phi$ describes a slow and small electric test 
field ${\cal E}(t)$ and, in addition, the vector potential $\f{A}$
contains a fast and strong external laser field inducing electronic
transitions. 
Performing an expansion into eigenfunctions of the undisturbed
operator $V-\na^2/(2m)$, we shall assume that only the three
lowest-lying eigenstates with the eigenfunctions $f_a(\f{r})$,
$f_b(\f{r})$, and $f_c(\f{r})$ and the eigenvalues 
$\omega_a<\omega_b<\omega_c$, respectively, are relevant
$\Psi(t,\f{r})=\psi_a(t)f_a(\f{r})e^{-i\omega_at}+
\psi_b(t)f_b(\f{r})e^{-i\omega_bt}+\psi_c(t)f_c(\f{r})e^{-i\omega_bt}$.
Note that we have used $\omega_b$ instead of $\omega_c$ in the last
term, which will be more convenient later on.

The frequency of the laser field is chosen such that it matches the
transition from the ground state $a$ to the first excited state $b$
-- but not from $b$ to $c$ or from $a$ to $c$.
Furthermore, we assume that the ground state $a$ is strongly localized
and hence does not couple significantly to the small and slow electric
test field ${\cal E}(t)$ -- whereas the states $b$ and $c$ are more
spread out and thus do couple.
Under these circumstances, the insertion of the above expansion into
the original Lagrangian in Eq.~(\ref{total}) and the application of
the rotating wave approximation together with the dipole approximation
yields the reduced Lagrangian
\bea
\label{RWA}
{\mathfrak L}_{\rm red}
&=&
i\psi_a^*\dot\psi_a+i\psi_b^*\dot\psi_b+i\psi_c^*\dot\psi_c
-\Delta\omega\,\psi_c^*\psi_c+
\nn&&+
(\kappa\,{\cal E}(t)\psi_b^*\psi_c+\Omega(t)\psi_a^*\psi_b+{\rm h.c.})
\,.
\ea
Here $\Delta\omega=\omega_c-\omega_b>0$ denotes the energy difference
between the states $a$ and $b$, the dipole moment $\kappa$ describes
the coupling to the small and slow electric test field ${\cal E}(t)$, 
and $\Omega(t)$ is the Rabi frequency of the laser.
Hence, the (approximated) equations of motion read
$i\dot\psi_a+\Omega(t)\psi_b=0$, 
$i\dot\psi_b+\kappa\,{\cal E}(t)\psi_c+\Omega(t)\psi_a=0$, and 
$i\dot\psi_c-\Delta\omega\,\psi_c+\kappa\,{\cal E}(t)\psi_b=0$.
Since the test field ${\cal E}(t)$ and thus the induced disturbance
$\psi_c$ are small (linear response), we may neglect the term 
$\kappa\,{\cal E}(t)\psi_c$ such that the dynamics for $\psi_a$ and
$\psi_b$ decouple and depend on $\Omega(t)$ only. 
In the adiabatic r\'egime, where the change of ${\cal E}(t)$ is much
slower than $\Delta\omega$, we can solve the remaining equation for
$\psi_c$ via $\psi_c(t)=\kappa{\cal E}(t)\psi_b(t)/\Delta\omega$,
which is equivalent to a mixing of the states $\psi_b$ and $\psi_c$ 
due to the disturbance ${\cal E}(t)$ as in stationary perturbation
theory. 
In the adiabatic limit, we may integrate out (i.e., average over)  
the microscopic degrees of freedom $\psi_{a,b,c}$ in the reduced
Lagrangian in Eq.~(\ref{RWA}) leading to the contribution to
the effective Lagrangian for the test field ${\cal E}(t)$
\bea
\label{eff-lag}
{\mathfrak L}_{\rm eff}
=|\psi_b(t)|^2\frac{\kappa^2}{\Delta\omega}\,{\cal E}^2(t)
\,.
\ea
Obviously this corresponds to a varying dielectric \cite{inductance} 
permittivity $\varepsilon(t)$ and thus capacitance
$C_n(t)$ which can be controlled (non-linear optics) by the laser field, 
i.e., its Rabi frequency $\Omega(t)$.
Note that, in contrast to resonant phenomena such as
electro-magnetically induced transparency, this effective Lagrangian
is valid for all adiabatic frequencies (i.e., far below $\Delta\omega$).

{\em Discussion}\quad
%
As we have observed in the previous considerations, it is possible 
(under the assumptions and approximations made) to increase the
capacitances in a controlled way by shining a laser beam on the
material without necessarily generating dissipation and noise etc. 
Note that a decoherence-free modulation of $\varepsilon(t)$ requires
the dipole moment for the transition $a \leftrightarrow b$ to be small
and the laser beam to contain a large number of photons -- 
such that it can be described classically as an external field, 
i.e., stimulated emission/absorption only.
In contrast, the spontaneous decay back to ground state $a$ is
potentially associated with dissipation, noise and decoherence -- 
and hence the life-time of the excited state $b$ should be much longer 
than all other time-scales relevant for the effective geometry. 

Fortunately, one can simulate a black hole horizon 
($c$ decreases $\leftrightarrow$ $\varepsilon$ increases) by
illuminating the material in its ground state $a$.
The white hole horizon occurs at the transition back to the ground
state $a$ and one has to make sure that the two horizons are far apart
such that the noise generated by the white hole horizon does not
propagate to the black hole horizon and prevents the detection of the
Hawking radiation. 
The dispersion relation following from Eq.~(\ref{discrete}) reads
(for $k_y=0$ \cite{compact})
\bea
\omega^2
=
\frac{4}{LC}\,\sin^2\left(\frac{k\Delta x}{2}\right)
\approx
c^2k^2-\frac{\Delta x^4}{12LC}\,k^4
\,,
\ea
and thus the group and phase velocities are 
(for moderate frequencies \cite{high}) 
less or equal to $c^2=\Delta x^2/(LC)$, i.e., we have a sub-luminal 
dispersion relation -- which is typical for a lattice. 
As a result, disturbances from the white hole horizon 
(with moderate frequencies) cannot propagate to the black hole horizon.

In order to answer the main question of whether it will be possible to
actually measure the Hawking effect one has to estimate the Hawking
temperature. 
According to Eq.~(\ref{Hawking}) the Hawking temperature is basically
determined by the characteristic time scale on which $c$ changes.
For our microscopic model, this switching time is related to the Rabi 
frequency $\Omega$, which must much smaller than the optical
frequencies $\omega_{a,b,c}$ for the rotating wave approximation to
apply. 
Nevertheless, with strong laser pulses, it is possible to pump a large 
number of electrons in a semi-conductor from the ground state into an 
excited state in 10-100 picoseconds -- whereas the (spontaneous) decay
back to the ground state can be much slower \cite{private}. 
(In a semi-conductor, the effective one-particle potential $V$ and
hence the dipole moment $\kappa$ and energy gap $\Delta\omega$ can
be very different from those of atomic transitions.)
In this case, the order of magnitude of the Hawking temperature could 
be ${10-100\;\rm mK}$, which is a really promising value since there
already exist amplifiers and detectors (for micro-waves) with a noise
temperature of order ${10\;\rm mK}$, see e.g., \cite{axion}.
Of course, one should cool down the apparatus below that temperature. 

One advantage of the present proposal is that it allows for large
velocities, say $c_0/10$-$c_0/100$ (in contrast to Bose-Einstein 
condensates \cite{bec}, for example, with a sound speed of order mm/s).
Furthermore, there are no walls in relative motion with respect to the
medium, which could lead to the Miles \cite{miles} instability 
(via momentum transfer).
With the above values, the thermal wave-length of the Hawking
radiation would be of order millimeter -- hence the minimum size
$\Delta x$ representing the analogue of the Planck length (knee
wave-length) should be smaller than that.
Consequently, it would be sufficient to illuminate a slab with a
thickness way below one millimeter by the laser.

The power of the radiation can be inferred from the (1+1 dimensional)
energy-momentum tensor (in the co-moving frame)
\bea
\frac{dE}{dt}=T^1_0=\frac{\pi}{12\hbar}\,
\left(k_{\rm B}T_{\rm Hawking}\right)^2
\,,
\ea
where we can get an order of magnitude between $10^{-14}$ and
$10^{-16}$ Watt, which vastly exceeds the threshold in \cite{axion}.
Owing to the Doppler effect (slow-down) the created power in the
laboratory frame is even larger.
The 1+1 dimensional character of the above expression reflects the
fact that we have excluded transverse modes via assumption
(\ref{delta}).
With appropriate generalizations (e.g., $\lambda\not\gg\Delta y$), 
we could also reproduce higher (e.g., 2+1) dimensional behavior
\cite{compact}.  

There are basically two major possibilities for the geometry of the
set-up -- a line or a circle.
In both cases one may accumulate energy over some time by building a
very long line (e.g., as a spiral) or using many revolutions in the
circle. 
In the latter scenario, however, the problem of how to deal with the
white hole horizon has to be solved.

The control of the space-time dependence of $c(t,x)$ can be achieved 
actively or passively:
An active control could be realized via sweeping the laser beam
externally (e.g., shining through holes in the lower capacitor plate), 
for example.
Alternatively (passive control), one could arrange optical fibers 
filling the capacitors -- either wound up around the capacitor plates 
or aligned along the wave-guide -- in a way such that the linear or 
non-linear (e.g., self-focusing) optical pulse propagation exactly 
generates the desired behavior $c(t,x)$.

The Hawking effect could be measured by connecting an amplifying
circuit (e.g., field effect transistors) to the end of the wave-guide
(an alternative method would be a bolometer) subject to impedance matching 
-- which can be achieved by manipulating $\mu/\varepsilon$, for example.
In case the measurement is not fast enough, that amplifier should be
disconnected (switched off) before the arrival of the horizon -- 
inducing additional noise etc. 
The thermal spectrum of the Hawking radiation could be determined with 
successive band-passes.

In summary, the present proposal for an experimental verification of the 
Hawking effect (in black hole analogues) appears to be just at the edge 
of the present experimental capabilities -- but not far beyond them 
(as is some other scenarios). 

{\em Acknowledgments}\quad
%
R.~S.~gratefully acknowledges fruitful discussions with D.~Bonn, D.~Broun, 
W.~Hardy, I.~Ozier, and G.~Ruoso as well as financial support by the
Emmy-Noether Programme of the German Research Foundation (DFG) under grant 
No.~SCHU 1557/1-1 and by the Humboldt foundation.  
Furthermore, this work was supported by the COSLAB Programme of the ESF, 
the CIAR, and the NSERC.

\addcontentsline{toc}{section}{References}


\begin{thebibliography}{99}
\bibitem{dual}
In the opposite case (constant capacitances and time-dependent inductances),
one can introduce dual potentials via $I_n=Cd\Lambda_n/dt$ and
$U_n=\Lambda_n-\Lambda_{n-1}$.
\bibitem{compact}
In view of the assumptions made, $A(t,x)$ does not depend on $y$,
i.e., the perpendicular wave-number $k_y$ vanishes.
%
Similar to an optical fiber based on total reflection, the set of
$k_y$ modes is discrete and the energy of the modes with $k_y>0$ 
is so large (for small $\Delta y$) that they can be neglected.
%
But taking into account the $y$-dependence, Eq.~(\ref{continuum})
generalizes to 
$\partial_t c_x^{-2} \partial_t A = \partial_x^2 A + \xi^2\partial_y^2 A$
where $\xi^2$ denotes the ratio of the (constant) inductances in the
$x$- and $y$-directions. 
%
After a simple re-scaling $y \to y\,\xi$, we reproduce
Eq.~(\ref{eff-metric}). 
%
Since Hawking radiation is basically a 1+1 dimensional effect with the
Hawking temperature being independent of the transversal dimensions
and conformal factors etc., the existence of the ''silent''
$y$-dimension does not alter the conclusions.
\bibitem{inductance}
In an analogous way one can generate a changing magnetic permeability
$\mu(t)$ and hence inductance $L_n(t)$.
\end{thebibliography}
\end{document}